\documentclass{article}
\usepackage{amsfonts}
\usepackage{bbm}
\usepackage{bm}
\usepackage{mathrsfs}
\usepackage{color}
\usepackage[left=2.00cm, right=2.00cm]{geometry}             
\usepackage{graphicx}
\usepackage{amssymb}
\usepackage{amsmath}
\usepackage{cite}
\usepackage{comment}
\usepackage{tikz}
\usetikzlibrary{matrix}

\usepackage[hyperindex=true,
          pdfstartview=FitH,
          bookmarksnumbered=true,
          bookmarksopen=true,
          citecolor=blue,
          linkcolor=blue,
          colorlinks=true,
          urlcolor=rossoCP3,
          pdfborder=001,
          unicode]{hyperref}

\definecolor{rossoCP3}{cmyk}{0,.88,.77,.40}

\newcommand{\blue}[1]{\textcolor{blue}{#1}}
\renewcommand{\blue}[1]{\textcolor{black}{#1}}

\newcommand{\re}{\mathrm{e}}

\usepackage{xcolor}  
\definecolor{mygreen}{RGB}{44,85,17}
\definecolor{myblue}{RGB}{34,31,217}
\definecolor{mybrown}{RGB}{194,164,113}
\definecolor{myred}{RGB}{255,66,56}
\definecolor{mypurple}{RGB}{200,36,176}

\allowdisplaybreaks

\begin{document}

\title{A general relativistic kinetic theory approach \\
to linear transport in generic hydrodynamic frame}

\author{Long Cui${}^1$, Xin Hao\thanks{Correspondig author.}{\hspace{.5em}}${}^2$, and Liu Zhao${}^1$
\vspace{1pt}\\
\small ${}^1$School of Physics, Nankai University, Tianjin 300071, China\\
\small ${}^2$School of Physics, Hebei Normal University, Shijiazhuang 050024, China\\
\small {email}: \href{mailto:cuilong@mail.nankai.edu.cn}{cuilong@mail.nankai.edu.cn},\\
\small \href{mailto:xhao@hebtu.edu.cn}{xhao@hebtu.edu.cn}
\small and \href{mailto:lzhao@nankai.edu.cn}{lzhao@nankai.edu.cn}}
\date{ }

\maketitle

\begin{abstract}
  In this study, we investigate the linear transport of neutral system
  within the framework of relativistic kinetic theory.
  Under the relaxation time approximation, we obtain an iterative solution to
  the relativistic Boltzmann equation in generic stationary spacetime. This solution
  provides a scheme to study non-equilibrium system order by order. Our calculations
  are performed in generic hydrodynamic frame, and the results can be reduced
  to a specific hydrodynamic frame by imposing constraints.
  As a specific example, we analytically calculated the covariant expressions of the particle
  flow and the energy momentum tensor up to the first order in relaxation time.
  Finally and most importantly, we present all 14 kinetic coefficients for a
  neutral system, which are verified to satisfy the Onsager reciprocal relation \blue{in a generic hydrodynamic frame}
  and guarantee a non-negative entropy production \blue{in the frame where the first order conservation laws are restored}.
\end{abstract}

\section{Introduction}

Relativistic kinetic theory sets up a powerful framework to study non-equilibrium
phenomena in curved spacetime. In general, at long wavelength and low frequency,
non-equilibrium systems can be excellently described by hydrodynamics which,
from a modern perspective, is an effective theory consisting of a gradient expansion
about the local equilibrium state. At linear level, the property of the fluid system is
totally determined by a set of phenomenological transport coefficients
that can be measured by experiment and can be determined by calculations in the
underlying microscopic theory. Although there are multiple approaches
with different domain of validity to calculate the transport coefficients,
in a curved spacetime background the relativistic kinetic
theory proves to be the most efficient and convenient option
with many important applications ranging from
stability theory {\cite{banach1994,Fajman:2017gaz,Rein:2023iic}},
astrophysics{\cite{Rioseco:2016jwc,Cieslik:2022wok}} to
cosmology {\cite{Weinberg:2003ur,Agon:2011mz,vereshchagin2017relativistic}}.

Historically, kinetic theory was developed to study thermodynamic behaviors
of classical gaseous systems. However, from quantum field theoretical point of view,
many thermodynamic properties of weakly coupled systems can be also
obtained using kinetic theory as an effective theory. One of the most well-known
examples is the effective kinetic theory for quark-gluon plasma
where, at high temperature or large density, the gauge coupling constant becomes
sufficiently small which allows for perturbative calculations
{\cite{Heinz:1983nx,Elze:1989un,Arnold:2007pg}}. Although
LHC experiment is the main motivation for interests
in quark-gluon plasma, such a state also occurs in the early universe and
the core of some neutron stars where the effect of spacetime curvature is
significant. Unfortunately, non-equilibrium thermodynamics in curved spacetime,
even for classical system, remains to some extent an open question.
In this direction, relativistic kinetic theory in curved spacetime
is expected to be relevant and applicable.

The earliest statistical description for the equilibrium state of relativistic gases
began with J\"{u}ttner {\cite{juttner1911}}, who extended
the Maxwell-Boltzmann distribution of equilibrium gases to the relativistic
case. Tolman and Ehrenfest
{\cite{Tolman:1930zza,Tolman:1930ona}} were the first to note that relativistic
gases can only be in equilibrium when the effects of the temperature gradient
and the gravitational field cancel out. Tauber and Weinberg
{\cite{Tauber:1961lbq}} had earlier formulated kinetic theory under general
relativity. Israel {\cite{israel1963}} established the non-equilibrium
distribution function and transport coefficients of relativistic gas by using
the Chapman-Enskog method. Lindquist {\cite{Lindquist:1966igj}} calculated the
transport equation of a gaseous system composed of zero rest mass particles in
spherically symmetric spacetime. Kremer {\cite{Kremer2012,Kremer2013}}
calculated the bulk viscosity and shear viscosity coefficients of gases in
Schwarzschild spacetime under the post-Newtonian approximation, obtaining
Fourier's laws for a single gas, and Fick's law for gas mixtures. However,
there is a scarce literature on the study of relativistic gases that deviate
from equilibrium in general curved spacetime in the framework of
relativistic kinetic theory.

Our goal is to calculate the kinetic coefficients for linear response in a covariant
formalism in a generic stationary spacetime. Here, by linear response we mean the
macroscopic phenomenon in the linear
order of generalized thermodynamic forces that drives
the system away from equilibrium, such as viscosity and heat conduction.
In the previous works
{\cite{Liu:2022wpu,Hao:2023xgq}}, the solution of the relativistic Boltzmann
equation was constructed through the gravito-electromagnetism analogy which
facilitate the discussion of particle and energy transport, but not the
viscosity phenomena. In the present work, we construct the solution of the
relativistic Boltzmann equation by use of an iterative procedure in terms of the
relaxation time, which allows for calculating all kinetic coefficients to any
order in relaxation time. Each of these kinetic coefficients can be expressed
as a function of temperature and chemical potential. Comparing to
previous literature, our formalism is simpler and more intuitive, and also applies
to the case of degenerate gases. For simplicity, this paper includes only
kinetic coefficients up to the first order in relaxation time, and focuses on
the gaseous system composed of massive neutral particles, which act as
a probe system in the background spacetime. In this framework, the computations
are fully analytical and clearly covariant.

\blue{
It is worth emphasizing that calculations are not carried out in a specific hydrodynamic
frame, then the transport coefficients with physical implications, can be derived
under appropriate scenarios.} On the other hand, physically relevant transport coefficients
can be also derived through the frame-independent combinations proposed by Kovtun \cite{Kovtun:2019hdm},
thereby rendering the choice of hydrodynamic frame non-essential in the study of kinetic coefficients.

The paper is structured as follows. In section \ref{Relativistic Boltzmann
equation and detailed balance}, we review the relativistic Boltzmann equation
and the detailed balance distribution. Section \ref{Deviation from the
detailed balance} discusses the deviation from the detailed balance and
iteratively solves the Boltzmann equation under the relaxation time
approximation. In section \ref{First order kinetic coefficients}, we use the
first order iterative solution to calculate all the transport coefficients up
to the first order in relaxation time. Section \ref{Onsager reciprocal relation}
verifies that the above result satisfies the Onsager reciprocal
relation and lead to a non-negative entropy production.

We adopt the metric signature $(-, +, \cdots, +)$, where
the dimension of space is $d$. The Greek letters $\mu, \nu, \cdots = 0, 1,
\cdots, d$ refer to spacetime indices and the \text{{\bfseries{}}}Latin
letters with hat $\hat{a}, \hat{b} = \hat{0}, \hat{1}, \cdots, \hat{d} ;
\enspace \hat{i}, \hat{j}, \cdots = \hat{1}, \hat{2}, \cdots, \hat{d}$ refer
to basis indices. In addition, quantities with an overbar (such as
$\bar{f}, \bar{T}, \bar{\mu}, \cdots$) indicate the values in detailed
balance.

\section{Relativistic Boltzmann equation and detailed
balance}\label{Relativistic Boltzmann equation and detailed balance}

Let us start with a brief review of the kinetic theory description
of relativistic fluids. The spacetime manifold $M$ is taken as general as
possible, and the corresponding tangent bundle is denoted as $T M$.
In relativistic kinetic theory, the one particle distribution function
(1PDF) $f$ is defined on the future mass shell bundle {\cite{Sarbach2012,Sarbach2013}}
\begin{equation}
  \Gamma_m^+ := \{ (x, p) \in T M|p^{\mu} p^{\nu} g_{\mu \nu} = - m^2 c^2\},
  \quad x \in M, p \in T_x M,
\end{equation}
where $p^{\mu}$ is the momentum which is timelike and future-directed,
and $m$ is the mass of particle.

The particle flow $N^{\mu}$, energy momentum tensor $T^{\mu \nu}$ and entropy
flow $S^{\mu}$ can be expressed in terms of the 1PDF $f (x, p)$ as
\begin{equation}
  N^{\mu} = c \int \ensuremath{\boldsymbol{\varpi}} p^{\mu} f, \quad T^{\mu
  \nu} = c \int \ensuremath{\boldsymbol{\varpi}} p^{\mu} p^{\nu} f,
  \label{N,T}
\end{equation}
\begin{equation}
  S^{\mu} = - k_B c \int \ensuremath{\boldsymbol{\varpi}} p^{\mu} f \left[
  \log \left( \frac{h^d f}{f^{\ast}   \mathfrak{g}} \right) -
  \frac{\mathfrak{g} \log f^{\ast}}{\varsigma h^d f} \right], \label{S}
\end{equation}
where $k_B$ is Boltzmann constant, $h$ is Planck constant, $\mathfrak{g}$ is
degree of degeneracy, $\varsigma = 0, 1, - 1$ denote the non-degenerate,
bosonic and fermionic cases respectively, $f^{\ast} = 1 + \varsigma
\mathfrak{g}^{- 1} h^d f$ and $\displaystyle\ensuremath{\boldsymbol{\varpi}} =
\frac{\sqrt{g}}{| p_0 |} (\mathrm{d} p)^d$ is the invariant volume element in
the momentum space (in which $g=|\det(g_{\mu\nu})|$).

It is customary to decompose $N^{\mu}, T^{\mu \nu}$ using the proper velocity
$U^{\mu}$ ($U^{\mu} U_{\mu} = - c^2$) of some prescribed observer $\mathcal O$,
\begin{eqnarray}
  N^{\mu} & = & n \, U^{\mu} + j^{\mu}, \nonumber\\
  T^{\mu \nu} & = & \frac{1}{c^2} \epsilon \, U^{\mu} U^{\nu} + \frac{1}{c^2}
  q^{\mu} U^{\nu} + \frac{1}{c^2} q^{\nu} U^{\mu} + (P + \Pi) \Delta^{\mu \nu}
  + \Pi^{\mu \nu}, \label{NT}
\end{eqnarray}
where $n$ is the particle number density,
$\varepsilon$ is the energy density, $j^{\mu}$ is the particle flux, $q^{\mu}$
is the energy flux, $P$ is the hydrostatic pressure, $\Pi$ is the dynamic
pressure, $\displaystyle\vspace{2pt}\Delta^{\mu \nu} = g^{\mu \nu} + \frac{1}{c^2} U^{\mu} U^{\nu}$
is the projection tensor, \blue{and the deviatoric stress tensor $\Pi^{\mu \nu}$ is trace-free}.
All the above mentioned hydrodynamic variables are measured by the observer $\mathcal O$, and they satisfy
the orthogonal relations $j^{\mu} U_{\mu} = q^{\mu} U_{\mu}  = \Pi^{\mu \nu}
U_{\mu} = 0$.

The evolution of the 1PDF is determined by the relativistic Boltzmann equation
\begin{equation}
  \mathcal{L}_{\mathcal{H}} f =\mathcal{C} (x, p),
\end{equation}
where
\begin{equation}
  \mathcal{L}_{\mathcal{H}} = p^{\mu}  \cfrac{\partial}{\partial x^{\mu}} -
  \Gamma^{\mu}_{\alpha \beta} p^{\alpha} p^{\beta}  \cfrac{\partial}{\partial
  p^{\mu}} \label{Liouville vector field}
\end{equation}
is the Liouville vector field which is tangent to both $T M$ and $\Gamma_m^+$, and
\begin{equation}
  \mathcal{C} (x, p) = \int \ensuremath{\boldsymbol{\varpi}}_2
  \ensuremath{\boldsymbol{\varpi}}_3  \ensuremath{\boldsymbol{\varpi}}_4  [W_x
  (p_3 + p_4 \mapsto p + p_2) f_3 f_4 f^{\ast} f_2^{\ast} - W_x (p + p_2
  \mapsto p_3 + p_4) f f_2 f_3^{\ast} f_4^{\ast}]
\end{equation}
is the collision integral {\cite{cercignani2002}} where $W_x$ is referred to
as the transition probability. Further, requiring that the collision process satisfies
microscopic reversibility $W_x (p_3 + p_4 \mapsto p + p_2) = W_x (p + p_2
\mapsto p_3 + p_4)$, the collision integral can be reduced to
\begin{equation}
  \mathcal{C} (x, p) = \int \ensuremath{\boldsymbol{\varpi}}_2
  \ensuremath{\boldsymbol{\varpi}}_3  \ensuremath{\boldsymbol{\varpi}}_4  [W_x
  (p_3 + p_4 \mapsto p + p_2)  (f_3 f_4 f^{\ast} f_2^{\ast} - f f_2
  f_3^{\ast} f_4^{\ast})] . \label{collision term}
\end{equation}

When the total entropy of the system reaches the maximum, the distribution
function no longer evolves (the collision integral vanishes), this
state is referred to as detailed balance and the corresponding 1PDF reads
\begin{equation}
  \bar{f} = \frac{\mathfrak{g}}{h^d}  \frac{1}{\re^{\bar{\alpha} -
  \bar{\mathcal{B}}_{\mu} p^{\mu}} - \varsigma} , \label{detailed balance}
\end{equation}
where the overbar emphasize that the variable takes value in detailed balance.
For a fluid composed of massive neutral particles in detailed balance, the relativistic Boltzmann equation
necessitates that $\bar{\alpha}$ is a constant scalar and $\bar{\mathcal{B}}^{\mu}$ is a
timelike Killing vector field, which in turn implies that the underlying spacetime
must be stationary. From a macroscopic perspective, this fluid configuration is referred to as
global equilibrium.

Let us substitute $\bar{\mathcal{B}}^{\mu} = \bar{\beta}
\bar{U}^{\mu}$ with $\bar{U}^{\mu}  \bar{U}_{\mu} =
- c^2$ into eq.~\eqref{detailed balance} and then insert the result
into eqs.~\eqref{N,T} and \eqref{S}.
Then the particle flow $\bar{N}^{\mu}$, the energy momentum tensor $\bar{T}^{\mu
\nu}$ and the entropy flow $\bar{S}^{\mu}$ in detailed balance follow,
\begin{eqnarray}
  \bar{N}^{\mu} & = & \bar{n} \, \bar{U}^{\mu} \nonumber\\
  & = & \frac{\mathfrak{g}}{h^d}  (m c)^d \mathcal{A}_{d - 1} \bar{J}_{d - 1,
  1}  \bar{U}^{\mu}, \label{detailed balance N} \\
  &  &  \nonumber\\
  \bar{T}^{\mu \nu} & = & \frac{1}{c^2}  \bar{\epsilon} \, \bar{U}^{\mu}
  \bar{U}^{\nu} + \bar{P}  \bar{\Delta}^{\mu \nu} \nonumber\\
  & = & \frac{\mathfrak{g}}{h^d} (m c)^{d + 1} c \mathcal{A}_{d - 1}
  \left( \bar{J}_{d - 1, 2}  \frac{1}{c^2}  \bar{U}^{\mu}  \bar{U}^{\nu} +
  \frac{1}{d}  \bar{J}_{d + 1, 0}  \bar{\Delta}^{\mu \nu} \right),
  \label{detailed balance T} \\
  &  &  \nonumber\\
  \bar{S}^{\mu} & = & \bar{s} \, \bar{U}^{\mu} \nonumber\\
  & = & k_B  \frac{\mathfrak{g}}{h^d}  (m c)^d \mathcal{A}_{d - 1} \left(
  \bar{\alpha}  \bar{J}_{d - 1, 1} + \bar{\beta} m c^2  \bar{J}_{d - 1, 2} +
  \frac{1}{d}  \bar{\beta} m c^2  \bar{J}_{d + 1, 0} \right)  \bar{U}^{\mu},
  \label{detailed balance S}
\end{eqnarray}
where we have introduced ${J}_{m, n}$ as a function of $({\alpha}, {\zeta} ={\beta} m c^2 )$
\begin{equation}
  {J}_{m, n} ({\alpha}, {\zeta}) \equiv \int_0^{\infty}
  \frac{\sinh^m \vartheta \cosh^n \vartheta}{\re^{{\alpha} + {\zeta} \cosh
  \vartheta} - \varsigma} \mathrm{d} \vartheta, \label{special function}
\end{equation}
and $\bar{J}_{m, n}$ is evaluated at $(\bar{\alpha}, \bar{\beta})$. Additionally, 
$\mathcal{A}_{d - 1}$ is the area of the $(d -
1)$-dimensional unit sphere $S^{d - 1}$, and $\displaystyle\vspace{4pt}
\bar{\Delta}^{\mu \nu} = g^{\mu
\nu} + \frac{1}{c^2}  \bar{U}^{\mu}  \bar{U}^{\nu}$. It
is easy to see that $\bar{N}^{\mu}$ is parallel to $\bar U^\mu$, and also
$\bar U^\mu$ is the unique normalized timelike eigenvector of $\bar{T}^{\mu
\nu}$. This means that the particle transport and energy transport directions are
parallel to $\bar{U}^{\mu}$. For this reason, $\bar{U}^{\mu}$ is interpreted
as the proper velocity of the fluid.

From eqs.~\eqref{detailed balance N}-\eqref{detailed balance S} we can extract the
particle number density $\bar{n}$, energy density $\bar{\epsilon}$, pressure
$\bar{P}$ and entropy density $\bar{s}$ measured by the {\em comoving observer}
$\bar{\mathcal{O}}$ (i.e. that with proper velocity $\bar{U}^\mu$) in detailed balance,
which are all function in $(\bar{\alpha}, \bar{\zeta})$:
\begin{eqnarray*}
  \bar{n} & = & - \frac{1}{c^2}  \bar{U}_{\mu}  \bar{N}^{\mu} =
  \frac{\mathfrak{g}}{h^d}  (m c)^d \mathcal{A}_{d - 1} \bar{J}_{d - 1, 1},\\
  \bar{\epsilon} & = & \frac{1}{c^2}  \bar{U}_{\mu}  \bar{U}_{\nu}
  \bar{T}^{\mu \nu} = \frac{\mathfrak{g}}{h^d} m^{d + 1} c^{d + 2}
  \mathcal{A}_{d - 1} \bar{J}_{d - 1, 2},\\
  \bar{P} & = & \frac{1}{d}  \bar{\Delta}_{\mu \nu}  \bar{T}^{\mu \nu} =
  \frac{\mathfrak{g}}{h^d} m^{d + 1} c^{d + 2}  \frac{\mathcal{A}_{d - 1}}{d}
  \bar{J}_{d + 1, 0},\\
  \bar{s} & = & - \frac{1}{c^2}  \bar{U}_{\mu}  \bar{S}^{\mu} = k_B
  \frac{\mathfrak{g}}{h^d}  (m c)^d \mathcal{A}_{d - 1} \left( \bar{\alpha}
  \bar{J}_{d - 1, 1} + \bar{\zeta} \bar{J}_{d - 1, 2} + \frac{1}{d}
  \bar{\zeta}  \bar{J}_{d + 1, 0} \right) .
\end{eqnarray*}
It's easy to verify that these four scalars satisfy the local Euler relation and the
Gibbs-Duhem relation
\begin{equation}
  \bar{s} = k_B \left( \bar{\alpha}  \bar{n} + \bar{\beta}  \bar{\epsilon} + \bar{\beta}
  \bar{P} \right),
\end{equation}
\begin{equation}
  - \bar{s} \, \mathrm{d} \left( \frac{1}{k_B  \bar{\beta}} \right) + \mathrm{d}
  \bar{P} + \bar{n} \, \mathrm{d} \left( \frac{\bar{\alpha}}{\bar{\beta}} \right)
  = 0,
\end{equation}
which implies that $\bar{\alpha}, \bar{\beta}$ have further
thermodynamic correspondence
\begin{equation}
  \bar{\alpha} = - \frac{\bar{\mu}}{k_B  \bar{T}}, \quad \bar{\beta} =
  \frac{1}{k_B  \bar{T}}, \label{thermodynamic correspondence}
\end{equation}
where $\bar{\mu}$ is the chemical potential and $\bar{T}$ is the temperature
in detailed balance.

Recall that $\bar{\alpha}$ is a constant scalar and $\bar{\mathcal{B}}^{\mu} =
\bar{\beta}  \bar{U}^{\mu}$ is Killing, i.e.
\begin{equation}
  \nabla_{\mu} \bar{\alpha} = 0, \quad \nabla_{(\mu} \bar{\mathcal{B}}_{\nu)}
  = 0. \label{killing equation}
\end{equation}
Substituting eq.~\eqref{thermodynamic correspondence} into eq.~\eqref{killing
equation} and decomposing the results into scalar, vector and tensor parts, we get the
following equations. For the scalar part, we have
\begin{equation}
  \bar{U}^{\mu} \nabla_{\mu} \bar{T} = 0, \quad \bar{U}^{\mu} \nabla_{\mu}
  \bar{\mu} = 0, \quad \nabla_{\mu} \bar{U}^{\mu} = 0, \label{scalar
  perturbation}
\end{equation}
which means that the temperature and chemical potential of the fluid in
detailed balance remain constant along the direction of motion, and
the fluid has no expansion. For the vector part, we have
\begin{equation}
  \nabla_{\mu} \bar{T} + \frac{\bar{T}}{c^2}  \bar{U}^{\nu} \nabla_{\nu}
  \bar{U}_{\mu} = 0, \quad \nabla_{\mu} \bar{\mu} + \frac{\bar{\mu}}{c^2}
  \bar{U}^{\nu} \nabla_{\nu} \bar{U}_{\mu} = 0, \label{vector perturbation}
\end{equation}
these two equations encode the well-known Tolman-Ehrenfest and the
Klein effects. For the tensor part,
\begin{equation}
  \bar{\Delta}^{\mu \rho} \bar{\Delta}^{\nu \sigma} \nabla_{(\rho } \bar{U}_{ \sigma)} -
  \frac{1}{d} \nabla_{\rho} \bar{U}^{\rho} \bar{\Delta}^{\mu \nu} = 0,
  \label{shear}
\end{equation}
which indicates that the fluid has no shear effect
in detailed balance.

In later calculations, we will not replace the variables $(\bar{\alpha},
\bar{\beta})$ by $(\bar{T}, \bar{\mu})$ for convenience.
In terms of $(\bar{\alpha}, \bar{\beta})$, eqs.~\eqref{scalar perturbation}
and \eqref{vector perturbation} can be written as
\begin{equation}
  \bar{U}^{\mu} \nabla_{\mu} \bar{\alpha} = 0, \quad \bar{U}^{\mu}
  \nabla_{\mu} \bar{\beta} = 0, \quad \nabla_{\mu} \bar{U}^{\mu} = 0,
  \label{detailed balance scalar}
\end{equation}
\begin{equation}
  \bar{\Delta}^{\mu \nu} \nabla_{\nu} \bar{\alpha} = 0, \quad \nabla_{\mu}
  \bar{\beta} - \frac{\bar{\beta}}{c^2}  \bar{U}^{\nu} \nabla_{\nu}
  \bar{U}_{\mu} = 0. \label{detailed balance vector}
\end{equation}
We will see in Section \ref{First order kinetic
coefficients} that the deviation from 0 of eqs.~
\eqref{shear}, \eqref{detailed balance scalar} and \eqref{detailed balance vector}
leads to correction terms for the particle flow and the energy momentum
tensor.

\section{Deviation from the detailed balance}\label{Deviation from the
detailed balance}

A subtle point that needs to be clarified is that for non-equilibrium systems
(even if the system is only slightly out of equilibrium), there is no prior
definition of the state parameters of the fluid (e.g.,
temperature, chemical potential and fluid velocity). One can
choose different ways to define the state parameters, but it is required
that the total particle flow and the energy momentum tensor are not
affected by different definitions, and that when the fluid returns to equilibrium,
the corresponding state parameters return to the same equilibrium state parameters.
Choosing a set of definitions of temperature,
chemical potential and fluid velocity is known as selecting a hydrodynamic frame
{\cite{Kovtun2012}}. In addition to the well-known Eckart frame and the Landau
frame, other hydrodynamic frames can also be selected.
Different hydrodynamic frames affect the division of
fluid orders without changing the physics itself.

For the near equilibrium fluid, we can reasonably use the local equilibrium
assumption to approximate the zeroth order 1PDF as
\begin{equation}
  f^{(0)} = \frac{\mathfrak{g}}{h^d}  \frac{1}{\re^{\alpha -\mathcal{B}_{\mu}
  p^{\mu}} - \varsigma}, \label{local equilibrium}
\end{equation}
and we still have the hydrodynamics implications
\begin{equation}
  \alpha = - \cfrac{\mu}{k_B T}, \quad \mathcal{B}^{\mu} = \beta U^{\mu} =
  \dfrac{1}{k_B T} U^{\mu},
\end{equation}
where $\mu$ is the local chemical potential, $T$ is the local
temperature and $U^{\mu}$ is still referred to as the velocity of the
fluid. However, $\alpha$ is no
longer constant and $\mathcal{B}^{\mu}$ needs not be Killing.
Therefor $\nabla_{\mu} \alpha$ and $\nabla_{(\mu} \mathcal{B}_{\nu)}$
provide a measure for how far the state of the fluid deviates from \blue{global equilibrium}.
The selected zeroth order 1PDF \eqref{local equilibrium} corresponds to a
perfect fluid, which is the zeroth order of the hydrodynamic derivative
expansion. When the system returns to equilibrium, $\alpha = \bar{\alpha},
\mathcal{B}^{\mu} = \bar{\mathcal{B}}^{\mu}$, eq.~\eqref{local equilibrium}
returns to the distribution function \eqref{detailed
balance}.

In view of kinetic theory, selecting a hydrodynamic frame means defining a zeroth
order 1PDF, which also means imposing constraints on $\alpha$ and $\mathcal{B}^{\mu}$.
In the following calculation, we just keep the form of $f^{(0)}$ without constraints on
$\alpha$ and $\mathcal{B}^{\mu}$. This approach is computationally flexible, while still
allowing for the imposition of constraints on the final result if necessary. These constraints can
limit the result to a specific hydrodynamic frame. When discussing the transport
coefficients in Section \ref{First order kinetic coefficients}, we will
return to the specific hydrodynamic frame by imposing constraints on
$\alpha, \beta$ and $U^{\mu}$.

The full Boltzmann equation is an extremely complicated integral-differential
equation, where the collision integral makes it difficult to solve
analytically. In this work, we adopt the standard relaxation time approximation
by replacing the collision integral with an Anderson-Witting-like collision
model
\begin{equation}
  \mathcal{C} (x, p) \simeq - \frac{\varepsilon}{c^2\tau}  (f - f^{(0)}) =
  \frac{U_{\mu} p^{\mu}}{c^2\tau}  (f - f^{(0)}),
\end{equation}
where $\varepsilon = - U_{\mu} p^{\mu}$ is the energy of a single particle
measured by the comoving observer $\mathcal O$, and $\tau$ is
the relaxation time which represents the time scale for the system to restore
balance.
{
It is crucial to highlight a pivotal aspect of the collision model. While the mathematical
form of the model is identical to that of the Anderson-Witting model
{\cite{anderson1974relativistic}}, its main distinction lies in the
hydrodynamic frame characterized by the choice of $U^\mu$.
In the standard Anderson-Witting model, adhering to the constraints
of the overall particle number conservation and energy-momentum tensor conservation, the hydrodynamic
frame in the model must be the Landau frame.  However, considering the ordering scheme of hydrodynamics,
the restriction of conservation laws is actually truncated at a specific hydrodynamic order,
and the Landau frame may not be a stable frame for the first-order theory \cite{Kovtun:2019hdm}.
In this sense, it is reasonable to first relax the constraints on $U^\mu$ in the Anderson-Witting type model,
and then restore the conservation laws by selecting an appropriate frame at specific order, which renders the
model more general and suitable for the study of linear response phenomena.

Next, we shall try to solve the Boltzmann equation using a standard iterative procedure.
To the first order in relaxation time, we have
\begin{equation}
  \mathcal{L}_H f^{(0)} = - \frac{\varepsilon}{c^2 \tau}  (f_{[1]} - f^{(0)}),
\end{equation}
where $f_{[1]} = f^{(0)} + f^{(1)}$ denotes the first order iterative
solution. Substituting eq.~\eqref{local equilibrium} into eq.~\eqref{Liouville vector field}, we
have
\begin{equation}
  \mathcal{L}_H f^{(0)} = (- p^{\mu} p^{\nu} \nabla_{\mu} \mathcal{B}_{\nu} +
  p^{\mu} \nabla_{\mu} \alpha)  \frac{\partial f^{(0)}}{\partial \alpha} .
\end{equation}
Thus the first order iterative solution reads
\begin{equation}
  f_{[1]} = f^{(0)} + f^{(1)} = f^{(0)} - \frac{c^2 \tau}{\varepsilon} (-
  p^{\mu} p^{\nu} \nabla_{\mu} \mathcal{B}_{\nu} + p^{\mu} \nabla_{\mu}
  \alpha)  \frac{\partial f^{(0)}}{\partial \alpha}. \label{first order
  recursive solution}
\end{equation}
It is clear that $\nabla_{(\mu} \mathcal{B}_{\nu)}$, $\nabla_{\mu}
\alpha$ represents the deviation of the system from the detailed balance.
In general, the iterative solution of order $n$ satisfies
\begin{equation}
  \mathcal{L}_H f_{[n - 1]} = - \frac{\varepsilon}{c^2 \tau}  (f_{[n]} -
  f^{(0)}), \label{iterative}
\end{equation}
Subsequently, the $n$-th order iterative solution $f_{[n]}$ can be expressed as
\begin{equation}
  f_{[n]} = \sum_{i = 0}^n  \left( - \frac{c^2
  \tau}{\varepsilon} \mathcal{L}_H \right)^i f^{(0)}. \label{iteration
  sequence}
\end{equation}
In the rest part of this work, we will focus on the first order iterative solution.

\section{First order kinetic coefficients}\label{First order kinetic
coefficients}

Using eqs.~\eqref{N,T} and \eqref{first order recursive solution}, we can
analytically calculate the particle flow and the energy momentum tensor (see
Appendix \ref{Calculation of particle flow and energy momentum
tensor} for details). At the zeroth order, we have
\begin{align}
  N^{(0) \mu} & = c \int \ensuremath{\boldsymbol{\varpi}} p^{\mu} f^{(0)} =
  n^{(0)} U^{\mu} \nonumber\\
  & = \frac{\mathfrak{g}}{h^d}  (m c)^d \mathcal{A}_{d - 1} J_{d - 1, 1} \label{N0}
  U^{\mu}, \\
  &  &  \nonumber\\
  T^{(0) \mu \nu} & = c \int \ensuremath{\boldsymbol{\varpi}} p^{\mu}
  p^{\nu} f^{(0)} = \frac{1}{c^2} \epsilon^{(0)} U^{\mu} U^{\nu} + P
  \Delta^{\mu \nu} \nonumber\\
  & = \frac{\mathfrak{g}}{h^d} m^{d + 1} c^{d + 2} \mathcal{A}_{d - 1}
  \left( J_{d - 1, 2}  \frac{1}{c^2} U^{\mu} U^{\nu} + \frac{1}{d} J_{d + 1,
  0} \Delta^{\mu \nu} \right).
\end{align}
Evidently, at the zeroth order, the constitutive equations match those of a system under detailed balance.
Therefore, the zeroth order is commonly referred to as local equilibrium. However, this state differs from
true equilibrium since the system is evolving at this stage, accounting for higher-order corrections.

At the first order, using the relation $\mathcal{B}^{\mu} = \beta
U^{\mu}$, we can simplify the results as
\begin{align}
  N^{(1) \mu} & = c \int \ensuremath{\boldsymbol{\varpi}} p^{\mu} f^{(1)}
  \nonumber\\
  & = n^{(1)} U^{\mu} + j^{(1) \mu}, \\
  &   \nonumber\\
  T^{(1) \mu \nu} & = c \int \ensuremath{\boldsymbol{\varpi}} p^{\mu}
  p^{\nu} f^{(1)} \nonumber\\
  & = \frac{1}{c^2} \epsilon^{(1)} U^{\mu} U^{\nu} + \frac{1}{c^2} q^{(1)
  \mu} U^{\nu} + \frac{1}{c^2} q^{(1) \nu} U^{\mu} + \Pi^{(1)} \Delta^{\mu
  \nu} + \Pi^{(1) \mu \nu},
\end{align}
where the first order hydrodynamic variables are
\begin{align}
  n^{(1)} & = - \frac{1}{c^2} U_{\mu} N^{(1) \mu} \nonumber\\
  & = - \tau \frac{\mathfrak{g}}{h^d} \mathcal{A}_{d - 1}  (m c)^d
  \left( \frac{\partial J_{d - 1, 1}}{\partial \alpha} U^{\nu}
  \nabla_{\nu} \alpha + m c^2  \frac{\partial J_{d - 1, 2}}{\partial \alpha}
  U^{\nu} \nabla_{\nu} \beta - m c^2  \frac{1}{d}  \frac{\partial J_{d + 1, 0}
  }{\partial \alpha} \beta \nabla_{\nu} U^{\nu} \right), \label{n1} \\
  &  \nonumber\\
  \epsilon^{(1)} & =  \frac{1}{c^2} U_{\mu} U_{\nu} T^{(1) \mu \nu}
  \nonumber\\
  & = - \tau \frac{\mathfrak{g}}{h^d} \mathcal{A}_{d - 1}  (m c)^{d + 1} c
  \left( \frac{\partial J_{d - 1, 2}}{\partial \alpha} U^{\rho}
  \nabla_{\rho} \alpha + m c^2  \frac{\partial J_{d - 1, 3} }{\partial \alpha}
  U^{\rho} \nabla_{\rho} \beta - m c^2  \frac{1}{d}  \frac{\partial J_{d + 1,
  1}}{\partial \alpha} \beta \nabla_{\rho} U^{\rho} \right), \\
  &   \nonumber\\
  \Pi^{(1)} & = \frac{1}{d} T^{(1) \mu \nu} \Delta_{\mu \nu} \nonumber\\
  & = - \tau \frac{\mathfrak{g}}{h^d} \mathcal{A}_{d - 1}  (m c)^{d + 1}
  \frac{c}{d} \left( \frac{\partial J_{d + 1, 0}}{\partial \alpha}
  U^{\rho} \nabla_{\rho} \alpha + m c^2  \frac{\partial J_{d + 1, 1}}{\partial
  \alpha} U^{\rho} \nabla_{\rho} \beta - m c^2  \frac{1}{d}  \frac{\partial
  J_{d + 3, - 1}}{\partial \alpha} \beta \nabla_{\rho} U^{\rho} \right), \\
  &   \nonumber\\
  j^{(1) \mu} & = \Delta^{\mu}_{\enspace \nu} N^{(1) \nu} \nonumber\\
  & = - \tau \frac{\mathfrak{g}}{h^d} \mathcal{A}_{d - 1}  (m c)^d
  \frac{c^2}{d} \left( \frac{\partial J_{d + 1, - 1}}{\partial \alpha}
  \nabla_{\nu} \alpha + m c^2  \frac{\partial J_{d + 1, 0} }{\partial \alpha}
  \left( \nabla_{\nu} \beta - \frac{\beta}{c^2} U^{\rho} \nabla_{\rho} U_{\nu}
  \right) \right) \Delta^{\mu \nu}, \\
  &   \nonumber\\
  q^{(1) \mu} & = - U_{\nu} T^{(1) \nu \sigma} \Delta^{\mu}_{\enspace
  \sigma} \nonumber\\
  & = - \tau \frac{\mathfrak{g}}{h^d} \mathcal{A}_{d - 1}  (m c)^{d + 1}
  \frac{c^3}{d} \left( \frac{\partial J_{d + 1, 0}}{\partial \alpha}
  \nabla_{\rho} \alpha + m c^2  \frac{\partial J_{d + 1, 1}}{\partial \alpha}
  \left( \nabla_{\rho} \beta - \frac{\beta}{c^2} U^{\sigma} \nabla_{\sigma}
  U_{\rho} \right) \right) \Delta^{\rho \mu}, \\
  &   \nonumber\\
  \Pi^{(1) \mu \nu} & = T^{(1) \rho \sigma} \Delta^{\mu}_{\enspace \rho}
  \Delta^{\nu}_{\enspace \sigma} - \frac{1}{d} T^{(1) \rho \sigma}
  \Delta_{\rho \sigma} \Delta^{\mu \nu} \nonumber\\
  & = \tau \frac{\mathfrak{g}}{h^d} \mathcal{A}_{d - 1}  (m c)^{d + 2}
  \frac{2c^2}{(d + 2) d}  \frac{\partial J_{d + 3, - 1}}{\partial \alpha} \beta
  \left( \Delta^{\rho \mu} \Delta^{\sigma \nu} \nabla_{(\rho } U_{ \sigma)} -
  \frac{1}{d} \nabla_{\rho} U^{\rho} \Delta^{\mu \nu} \right) . \label{pai1}
\end{align}

\

From the tensor part $\displaystyle\Pi^{(1) \mu \nu} = - \eta \left( \Delta^{\rho \mu}
\Delta^{\sigma \nu} \nabla_{(\rho } U_{ \sigma)} - \frac{1}{d} \nabla_{\rho}
U^{\rho} \Delta^{\mu \nu} \right)$, we can read off the shear viscosity coefficient
\begin{equation}
  \eta = - c^2 \tau \frac{\mathfrak{g}}{h^d} \mathcal{A}_{d - 1}  (m c)^{d +
  2}  \frac{2}{(d + 2) d}  \frac{\partial J_{d + 3, - 1}}{\partial \alpha}
  \beta .
\end{equation}
According to the definition of the special function $J_{m, n}$, we can easily conclude that $\eta > 0$ and $\eta$ increases with
temperature.

Now, introducing a new variable $u = \zeta(\cosh \vartheta - 1)$, we can rewrite the
integral $J_{m, n}$ as
\[ J_{m, n} = \int_0^{\infty} \frac{\left( \frac{u^2}{\zeta^2} + 2 \frac{u}{\zeta}
   \right)^{\frac{m - 1}{2}} \left( \frac{u}{\zeta} + 1 \right)^n}{\zeta
   (\re^{\alpha + \zeta + u} - \varsigma)} \mathrm{d} u. \]
In the high temperature limit, $\beta \rightarrow 0$,
$\zeta = \beta m c^2 \rightarrow 0$, we have
\begin{eqnarray}
  J_{m, n} & \rightarrow & \left( \frac{1}{\zeta} \right)^{m + n}
  \int_0^{\infty} \frac{u^{m + n - 1}}{\re^{\alpha + u} - \varsigma} \mathrm{d}
  u = \left( \frac{1}{\zeta} \right)^{m + n} \frac{1}{\varsigma} \Gamma (m + n)
  \ensuremath{\operatorname{Li}}_{m + n} (\re^{- \alpha} \varsigma), \\
  \frac{\partial J_{m, n}}{\partial \alpha} & \rightarrow & - \left(
  \frac{1}{\zeta} \right)^{m + n} \frac{1}{\varsigma} \Gamma (m + n)
  \ensuremath{\operatorname{Li}}_{m + n - 1} (\re^{- \alpha} \varsigma) .
\end{eqnarray}
It is easy to see that the shear viscosity coefficient $\eta$ diverges in the
high temperature limit due to the asymptotic behavior
$\cfrac{\partial J_{d + 3, - 1}}{\partial
\alpha} \beta \sim \left( \cfrac{1}{\zeta} \right)^{d + 1}$.

The vector part $\{ j^{(1) \mu}, q^{(1) \mu} \}$ has been investigated in the
previous works {\cite{Liu:2022wpu,Hao:2023xgq}}, and it has been shown that the Onsager
reciprocal relation and Wiedemann-Franz law hold. Nevertheless,
it is meaningful to show again that the calculation of the heat
conductivity coefficient.

{\blue{The heat flow is defined to be the internal energy flow in the absence of a net particle
number flow i.e. $j^{(1)} = 0$, which gives a constraint} between $\Delta^{\mu \nu}
\nabla_{\nu} \alpha$ and $\Delta^{\mu \nu} \left( \nabla_{\nu} \beta -
\dfrac{\beta}{c^2} U^{\rho} \nabla_{\rho} U_{\nu} \right)$:
\begin{equation}
  \Delta^{\mu \nu} \left( \frac{\partial J_{d + 1, - 1}}{\partial \alpha}
  \nabla_{\nu} \alpha + m c^2  \frac{\partial J_{d + 1, 0} }{\partial \alpha}
  \left( \nabla_{\nu} \beta - \frac{\beta}{c^2} U^{\rho} \nabla_{\rho} U_{\nu}
  \right) \right) = 0.
\end{equation}
In such case, the energy flow is simply the heat flow. We can eliminate
$\Delta^{\mu \nu} \nabla_{\nu} \alpha$ and making use of the relation $T = \dfrac{1}{k_B \beta}$
to express the heat flow as
\begin{equation}
  q^{(1) \mu} = - \kappa \left( \nabla_{\rho} T + \frac{T}{c^2} U^{\sigma}
  \nabla_{\sigma} U_{\rho} \right) \Delta^{\rho \mu},
\end{equation}
where the heat conductivity coefficient reads
\begin{equation}
  \kappa = - \tau \frac{\mathfrak{g}}{h^d} \mathcal{A}_{d - 1}  (m c)^{d + 2}
  \frac{c^4}{d}  \frac{1}{k_B T^2} \left( - \frac{\left( \dfrac{\partial J_{d
  + 1, 0}}{\partial \alpha} \right)^2}{\dfrac{\partial J_{d + 1, -
  1}}{\partial \alpha}} + \frac{\partial J_{d + 1, 1}}{\partial \alpha}
  \right) .
\end{equation}
It is east to verify that $\kappa > 0$ and $\kappa$ increases with
temperature. In the high temperature limit, $\kappa \sim \left( \cfrac{1}{\zeta}
\right)^d$.

\blue{
To investigate bulk viscosity, the relevant physical scenario is typically defined by
$n^{(1)} = 0, \epsilon^{(1)} = 0$, which is consistent with the definition of bulk viscosity
in the non-relativistic framework. Subsequently, the constraint equations can be expressed as}
\begin{align}
  \frac{\partial J_{d - 1, 1}}{\partial \alpha} U^{\nu} \nabla_{\nu} \alpha +
  m c^2  \frac{\partial J_{d - 1, 2}}{\partial \alpha} U^{\nu} \nabla_{\nu}
  \beta - m c^2  \frac{1}{d}  \frac{\partial J_{d + 1, 0} }{\partial \alpha}
  \beta \nabla_{\nu} U^{\nu} & = 0, \\
  \frac{\partial J_{d - 1, 2}}{\partial \alpha} U^{\rho} \nabla_{\rho} \alpha
  + m c^2  \frac{\partial J_{d - 1, 3} }{\partial \alpha} U^{\rho}
  \nabla_{\rho} \beta - m c^2  \frac{1}{d}  \frac{\partial J_{d + 1,
  1}}{\partial \alpha} \beta \nabla_{\rho} U^{\rho} & = 0.
\end{align}
From these two equations, we can express $U^{\nu} \nabla_{\nu} \alpha$ and
$U^{\nu} \nabla_{\nu} \beta$ in terms of $\nabla_{\nu} U^{\nu}$, so that
$\Pi^{(1)}$ can be written as
\begin{equation}
  \Pi^{(1)} = - \xi \nabla_{\rho} U^{\rho},
\end{equation}
where the bulk viscosity coefficient $\xi$ reads
\begin{align}
  \xi & = c^2 \tau \frac{\mathfrak{g}}{h^d}  \frac{\mathcal{A}_{d -
  1}}{d^2}  (m c)^{d + 2} \beta  \nonumber\\
  & ~~~~ \times\left( \frac{ \left( \dfrac{\partial J_{d + 1, 0} }{\partial \alpha}
  \right)^2  \dfrac{\partial J_{d - 1, 3} }{\partial \alpha} + \left(
  \dfrac{\partial J_{d + 1, 1}}{\partial \alpha} \right)^2  \dfrac{\partial
  J_{d - 1, 1}}{\partial \alpha} - 2 \dfrac{\partial J_{d - 1, 2}}{\partial
  \alpha}  \dfrac{\partial J_{d + 1, 1}}{\partial \alpha}  \dfrac{\partial
  J_{d + 1, 0}}{\partial \alpha}}{- \left( \dfrac{\partial J_{d - 1,
  2}}{\partial \alpha} \right)^2 + \dfrac{\partial J_{d - 1, 1}}{\partial
  \alpha}  \dfrac{\partial J_{d - 1, 3} }{\partial \alpha} } - \frac{\partial
  J_{d + 3, - 1}}{\partial \alpha} \right),
\end{align}
which, in the high temperature limit, behaves as
$\xi \sim \left( \cfrac{1}{\zeta} \right)^{d -3}$. The different asymptotic behaviors
for the shear viscosity coefficient $\eta$ and the bulk viscosity coefficient $\xi$
indicate that these two viscosities are of different order of magnitude in nature.
The above results extends Kremer's works {\cite{Kremer2012}} on
shear and bulk viscosity to generic stationary backgrounds.

\section{Onsager reciprocal relation and local entropy production}
\label{Onsager reciprocal relation}

\blue{The results of the previous section are obtained in general hydrodynamic frame.
According to the expressions of the first order quantity of hydrodynamics
\eqref{n1}-\eqref{pai1}, we can simply write the constitutive relations of
the first order hydrodynamic variables as
\begin{align}
  n^{(1)} & = \nu_1 U^{\nu} \nabla_{\nu} \alpha + \nu_2 U^{\nu} \nabla_{\nu}
  \beta - \nu_3 \beta \nabla_{\nu} U^{\nu}, \nonumber\\
  \epsilon^{(1)} & = \rho_1 U^{\nu} \nabla_{\nu} \alpha + \rho_2 U^{\nu}
  \nabla_{\nu} \beta - \rho_3 \beta \nabla_{\nu} U^{\nu}, \nonumber\\
  \Pi^{(1)} & = \pi_1 U^{\nu} \nabla_{\nu} \alpha + \pi_2 U^{\nu}
  \nabla_{\nu} \beta - \pi_3 \beta \nabla_{\nu} U^{\nu}, \nonumber\\
  j^{(1) \mu} & = \sigma_1 \Delta^{\mu \nu} \nabla_{\nu} \alpha + \sigma_2
  \Delta^{\mu \nu} \left( \nabla_{\nu} \beta - \frac{\beta}{c^2} U^{\rho}
  \nabla_{\rho} U_{\nu} \right), \label{constitutive relation} \\
  q^{(1) \mu} & = \kappa_1 \Delta^{\mu \nu} \nabla_{\nu} \alpha + \kappa_2
  \Delta^{\mu \nu} \left( \nabla_{\nu} \beta - \frac{\beta}{c^2} U^{\rho}
  \nabla_{\rho} U_{\nu} \right), \nonumber\\
  \Pi^{(1) \mu \nu} & = - \eta \left( \Delta^{\rho \mu} \Delta^{\sigma \nu}
  \nabla_{(\rho } U_{ \sigma)} - \frac{1}{d} \nabla_{\rho} U^{\rho}
  \Delta^{\mu \nu} \right), \nonumber
\end{align}
where all the kinetic coefficients $\{ \nu_i, \sigma_i, \rho_i, \kappa_i,
\pi_i, \eta \}$ are calculated in eqs.~\eqref{n1}-\eqref{pai1} and the
generalized force can be defined as follow
\begin{align}
  \ensuremath{\boldsymbol{F}}  \equiv &~ \biggl[ U^{\mu}
  \nabla_{\mu} \alpha, U^{\mu} \nabla_{\mu} \beta, - \beta \nabla_{\mu}
  U^{\mu}, \Delta^{\mu \nu} \nabla_{\nu} \alpha, \Delta^{\mu \rho} \left(
  \nabla_{\rho} \beta - \frac{\beta}{c^2} U^{\sigma} \nabla_{\sigma} U_{\rho}
  \right), - \beta \left( \Delta^{\mu \rho} \Delta^{\nu \sigma} \nabla_{(\rho}
  U_{\sigma)} - \frac{1}{d} \nabla_{\rho} U^{\rho} \Delta^{\mu \nu} \right)
  \biggl], \label{GF}
\end{align}
which vanishes in the detailed balance state (c.f. eqs.~\eqref{detailed balance
scalar},\eqref{detailed balance vector} and \eqref{shear}). Notably, the coefficient
matrix is not only symmetric in the vector part of the constitutive relation, but also symmetric
in the scalar part, i.e.
\begin{equation}
 \left( \begin{array}{cc}
     \sigma_1 & \sigma_2\\
     \kappa_1 & \kappa_2
   \end{array} \right), \quad \left( \begin{array}{ccc}
     \nu_1 & \nu_2 & \nu_3\\
     \rho_1 & \rho_2 & \rho_3\\
     \pi_1 & \pi_2 & \pi_3
   \end{array} \right).
\end{equation}
Thus, we have verified a more complete Onsager reciprocity relation
in general hydrodynamic frame.}

\blue{To demonstrate the positivity of entropy production, the conservation laws must be restored
to the first order in relaxation time, which can be achieved by selecting an appropriate frame.
We will first outline the process for determining the
hydrodynamic frame.} According to the following identity:
\begin{align*}
\nabla_\mu \int\bm\varpi p^\mu  \psi f= \int \bm\varpi f \mathcal L_\mathcal{H} \psi
+  \int \bm\varpi \psi \mathcal L_\mathcal{H} f,
\end{align*}
by considering $\psi = c$ and $\psi = c\,p_\mu \xi^\mu$ where $\xi^\mu$ represents an arbitrary
vector field independent of $p$, we obtain
\begin{align}
\nabla_\mu N^\mu= c \int \bm\varpi  \mathcal L_\mathcal{H} f, \qquad
\nabla_\mu T^{\mu \nu}= c \int \bm\varpi \, p^\nu \mathcal L_\mathcal{H} f.
\label{FCE}
\end{align}
Subsequently, we substitute the Anderson-Witting type collision model into the above equation
which yields
\begin{align*}
\nabla_\mu N^\mu= \frac{1}{c^2 \tau} U_\mu \delta N^\mu = -\frac{1}{\tau} \delta n, \qquad
\nabla_\mu T^{\mu \nu}= \frac{1}{c^2 \tau} U_\mu \delta T^{\mu \nu}
= -\frac{1}{c^2 \tau}\left(\delta \epsilon \, U^\nu + q^\nu \right).
\end{align*}
where
\begin{align*}
\delta N^{\mu} = c \int \bm{\varpi} p^{\mu} (f - f^{(0)}), \qquad \delta
T^{\mu \nu} = c \int \bm{\varpi} p^{\mu} p^{\nu} (f - f^{(0)}),
\end{align*}
and the corresponding hydrodynamic decompositions are
\begin{align*}
\delta N^{\mu} = \delta n U^{\mu} + j^{\mu}, \qquad \delta T^{\mu \nu} =
\frac{1}{c^2} \delta \epsilon U^{\mu} U^{\nu} + \frac{2}{c^2} q^{(\mu}
U^{\nu)} + \Pi \Delta^{\mu \nu} + \Pi^{\mu \nu}.
\end{align*}
The overall conservation laws require that
$ \delta n = 0, \delta \epsilon = 0, q^\mu = 0 $, corresponding to the Landau frame in
hydrodynamics. Moreover, equation \eqref{FCE} can be used to discuss the conservation law order by order.
\blue{Expand $ \delta n, \delta \epsilon$ and $ q^\mu $ in powers of relaxation time
\begin{align*}
\delta n &= n^{(1)} + n^{(2)} + \cdots, \\
\delta \epsilon &= \epsilon^{(1)} + \epsilon^{(2)} + \cdots, \\
q^{\mu} &= q^{(1)\mu} + q^{(2)\mu} + \cdots.
\end{align*}
Up to the leading order, the following condition
\[
n^{(1)}=0,  ~~~~\epsilon^{(1)}=0, ~~~~q^{(1)\mu}=0,
\]
correspond to the zero order conservation equations $\nabla_\mu N^{(0)\mu}=0, ~ \nabla_\mu T^{(0)\mu\nu}=0,$
on the hydrodynamic side. For instance, it can be verified that $\delta n^{(1)}=0$ and $\nabla_\mu N^{(0)\mu}=0$
lead to the same equation by using Eqs. \eqref{N0}, \eqref{n1} and the properties of $J_{n,l}(\alpha,\zeta)$
\[
\frac{\partial J_{n, l}}{\partial \alpha} = - \frac{n - 1}{\zeta} J_{n - 2, l
+ 1} - \frac{l}{\zeta} J_{n, l - 1}, \quad \frac{\partial J_{n, l}}{\partial
\alpha} = \frac{\partial J_{n, l - 1}}{\partial \zeta}.
\]
Alternatively, one can impose conservation conditions
to the first order of relaxation time while maintaining $n^{(1)}$, $\epsilon^{(1)}$ and $q^{(1)\mu}$ nonzero.
In this case, a second-order iterative solution becomes essential. When the quadratic terms of
thermodynamic forces are neglected, detailed calculations reveal that
\[
\nabla_\mu \left(N^{(0)\mu} + N^{(1)\mu}\right) = -\frac{1}{\tau}\left(n^{(1)}+ n^{(2)}\right),
\]
\[
\nabla_\mu \left(T^{(0)\mu\nu} + T^{(1)\mu\nu}\right) = -\frac{1}{c^2\tau}\left[
\left(\epsilon^{(1)}+ \epsilon^{(2)}\right)U^\nu
+ \left(q^{(1)\nu} + q^{(2)\nu}\right)\right].
\]
Therefore, if the conservation equations are required to the first order of relaxation time,
the corresponding constraints are:
\[
n^{(1)} + n^{(2)} = 0, \qquad
\epsilon^{(1)} + \epsilon^{(2)} = 0, \qquad
q^{(1)\mu } + q^{(2)\mu } = 0.
\]
which form a closed system of equations involving temperature, chemical potential and fluid velocity,
thereby determining the hydrodynamic frame.}

According to the definition of entropy flow $S^{\mu}$ \eqref{S}, up to the
zeroth order in relaxation time, we can verify the covariant Euler
relation and the covariant Gibbs-Duhem relation
\begin{align}
  S^{(0) \mu}  = &~ - k_B c \int \ensuremath{\boldsymbol{\varpi}} p^{\mu}
  f^{(0)}  \left[ \log \left( \frac{h^d f^{(0)}}{{f^{(0)}}^{\ast}
  \mathfrak{g}} \right) - \frac{{\mathfrak{g} \log f^{(0)}}^{\ast}}{\varsigma
  h^d f^{(0)}} \right] \nonumber\\
   = &~ k_B (\Xi^{\mu} -\mathcal{B}_{\nu} T^{(0) \mu \nu} + \alpha N^{(0)
  \mu}),
\end{align}
\begin{equation}
  \mathrm{d} \Xi^{\mu} = T^{(0) \mu \nu} \mathrm{d} \mathcal{B}_{\nu} - N^{(0)
  \mu} \mathrm{d} \alpha,
\end{equation}
where
\begin{equation}
  \Xi^{\mu} = P\mathcal{B}^{\mu} = c \int \ensuremath{\boldsymbol{\varpi}}
  p^{\mu}  \frac{\mathfrak{g}}{\varsigma h^d} {\log \bigg(1+\frac{\varsigma h^d}{\mathfrak{g}} f\bigg)}.
\end{equation}
Considering that in the first order, the deviation $f^{(1)}$ from the local
equilibrium is a small correction, then
\begin{align}
  S^{(1) \mu}  = &~ - k_B c \int \ensuremath{\boldsymbol{\varpi}} p^{\mu}
  f^{(1)} \log \frac{f^{(0)}}{\mathfrak{g}/ h^d + \varsigma f^{(0)} }
  \nonumber\\
   = &~ k_B c \int \ensuremath{\boldsymbol{\varpi}} p^{\mu} f^{(1)}  (\alpha
  -\mathcal{B}_{\nu} p^{\nu}) \nonumber\\
   = &~ k_B  (\alpha N^{(1) \mu} -\mathcal{B}_{\nu} T^{(1) \mu \nu}) .
\end{align}
Therefore, up to the first order, we can conclude that
\begin{equation}
  S^{\mu} = S^{(0) \mu} + S^{(1) \mu} = k_B (P\mathcal{B}^{\mu}
  -\mathcal{B}_{\nu} T^{\mu \nu} + \alpha N^{\mu}) .
\end{equation}

Up to the first order of relaxation time, using the conservation equation $\nabla_{\mu} N^{\mu} =
0$, $\nabla_{\mu} T^{\mu \nu} = 0$ and the corollary derived from the
Gibbs-Duhem relation
\begin{equation}
  \nabla_{\mu} \Xi^{\mu} = T^{(0) \mu \nu} \nabla_{\mu} \mathcal{B}_{\nu} -
  N^{(0) \mu} \nabla_{\mu} \alpha,
\end{equation}
we can obtain
\begin{align}
  \frac{1}{k_B} \nabla_{\mu} S^{\mu}  = &~ - T^{(1) \mu \nu} \nabla_{\mu}
  \mathcal{B}_{\nu} + N^{(1) \mu} \nabla_{\mu} \alpha \nonumber\\
   = &~ n^{(1)} U^{\mu} \nabla_{\mu} \alpha
  + \epsilon^{(1)} U^{\mu} \nabla_{\mu} \beta - \Pi^{(1)} \beta \nabla_{\mu}
  U^{\mu} \nonumber\\
    &~ + j^{(1) \mu} \Delta_{\mu}^{\enspace \nu} \nabla_{\nu} \alpha + q^{(1)
  \mu} \Delta_{\mu}^{\enspace \rho} \left( \nabla_{\rho} \beta -
  \frac{\beta}{c^2} U^{\sigma} \nabla_{\sigma} U_{\rho} \right) \nonumber\\
    &~ - \Pi^{(1) \mu \nu} \beta \left( \Delta_{\mu}^{\enspace \rho}
  \Delta_{\nu}^{\enspace \sigma} \nabla_{(\rho} U_{\sigma)} - \frac{1}{d}
  \nabla_{\rho} U^{\rho} \Delta_{\mu \nu} \right) . \label{entropy generation}
\end{align}
\blue{Evidently, the local entropy production is driven by the generalized force $\boldsymbol{F}$
presented in equation \eqref{GF}.} It is worth noting that in addition to the vector part $\{ j^{(1) \mu}, q^{(1) \mu}\}$,
the entropy production also contains contributions from the scalar part
$\{ n^{(1)}, \epsilon^{(1)}, \Pi^{(1)} \}$ and the tensor part $\{ \Pi^{(1)
\mu \nu} \}$.

We are now in a position to investigate whether the approximate solution
\eqref{first order recursive solution} ensure a non-negative entropy
production at the first order in relaxation time. Substituting
eq.~\eqref{constitutive relation} into eq.~\eqref{entropy generation} we can obtain a
quadratic form
\begin{equation}
   \frac{1}{k_B} \nabla_{\mu} S^{\mu} = \ensuremath{\boldsymbol{F}} \left(
   \begin{array}{ccc|cc|c}
     \nu_1 & (\nu_2 + \rho_1) / 2 & (\nu_3 + \pi_1) / 2 & 0 & 0 & 0\\
     (\nu_2 + \rho_1) / 2 & \rho_2 & (\rho_3 + \pi_2) / 2 & 0 & 0 & 0\\
     (\nu_3 + \pi_1) / 2 & (\rho_3 + \pi_2) / 2 & \pi_3 & 0 & 0 & 0\\
     \hline
     0 & 0 & 0 & \sigma_1 & (\sigma_2 + \kappa_1) / 2 & 0\\
     0 & 0 & 0 & (\sigma_2 + \kappa_1) / 2 & \kappa_2 & 0\\
     \hline
     0 & 0 & 0 & 0 & 0 & \eta / \beta
   \end{array} \right) \ensuremath{\boldsymbol{F}}^T,
   \label{blockm}
\end{equation}
and the non-negative entropy production means that all eigenvalues of the
quadratic form are non-negative.

For the tensor part $\{ \Pi^{(1) \mu \nu} \}$, the only transport coefficient
$\eta > 0$ ensures that the eigenvalue of the tensor part is non-negative.

For the vector part $\{ j^{(1) \mu}, q^{(1) \mu} \}$, the condition that the
quadratic form is non-negative can be written as $\sigma_1 \geqslant 0, 4
\sigma_1 \kappa_2 - (\sigma_2 + \kappa_1)^2 \geqslant 0$. Substituting the
value of the transport coefficient, it is easy to verify that both conditions
are satisfied
\begin{align}
  \sigma_1  = &~ - c^2 \tau \frac{\mathfrak{g}}{h^d}  \frac{\mathcal{A}_{d -
  1}}{d}  (m c)^d  \frac{\partial J_{d + 1, - 1}}{\partial \alpha} > 0,\\
  4 \sigma_1 \kappa_2 - (\sigma_2 + \kappa_1)^2  = &~ 4 \left( \tau
  \frac{\mathfrak{g}}{h^d}  (m c)^{d + 1} \mathcal{A}_{d - 1}  \frac{c^3}{d}
  \right)^2 \left[ \frac{\partial J_{d + 1, - 1}}{\partial \alpha}
  \frac{\partial J_{d + 1, 1}}{\partial \alpha} - \left( \frac{\partial J_{d +
  1, 0}}{\partial \alpha} \right)^2 \right] > 0.
\end{align}

\begin{figure}[thb]
  \centerline{\includegraphics[width=.95\textwidth]{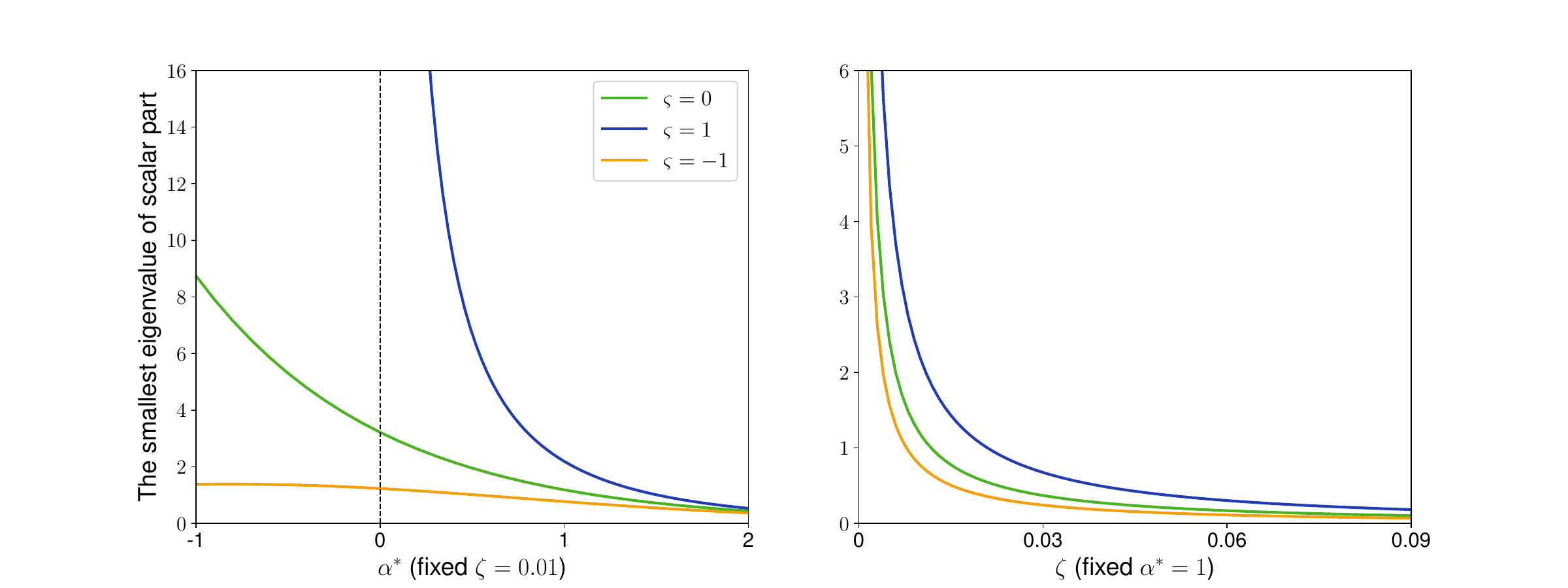}}
  \caption{The smallest eigenvalue of the dimensionless coefficient matrix for
  the scalar part. The left plot depicts the smallest eigenvalue as a function
  of $\alpha^{\ast} = \alpha + \zeta$ at fixed $\zeta = \beta m c^2 = 0.01$. The
  right plot depicts the smallest eigenvalue as a function of $\zeta$ at fixed
  $\alpha^{\ast} = 1$. The curve with $\varsigma = 1$ in the left plot does
  not cross the vertical line at $\alpha^{\ast} = 0$ because the chemical
  potential $\mu$ of relativistic Bose gas is not greater than $m c^2$.\label{scalareig}}
\end{figure}

For the scalar parts $\{ n^{(1)}, \epsilon^{(1)}, \Pi^{(1)} \}$
(which correspond to the upper-left block of the matrix in eq.~\eqref{blockm}),
analytical determination of the eigenvalues proves intractable.
Nevertheless, numerical analysis reveals no eigenvalues below zero.
We illustrate this by plotting the smallest eigenvalue of the dimensionless
coefficient matrix for the scalar parts, as shown in Figure~\ref{scalareig}.
Consequently, enforcing the conservation equation for the first-order fluid,
$\nabla_{\mu} S^{\mu} \geqslant 0$ is automatically satisfied.

\section{Conclusions and remarks}

In this work, we have iteratively solved the relativistic Boltzmann equation
under the relaxation time approximation in generic stationary spacetime,
calculated the first order hydrodynamic variables using the first order solution,
and analyzed the first order kinetic coefficients in general hydrodynamic frame.
\blue{Moreover, utilizing these kinetic coefficients, we derived several physically meaningful transport coefficients,
including shear viscosity, bulk viscosity, and heat conductivity. We also examined the asymptotic behavior of
these transport coefficients in the high temperature limit.}

Additionally, given the important
role of Onsager reciprocal relation in conventional non-equilibrium statistical
physics {\cite{Onsager:1931jfa,Onsager:1931kxm}}, we further analyzed these kinetic
coefficients. Our calculations show that, up to the first order of the relaxation time
the kinetic coefficients satisfy a more generalized
Onsager reciprocal relation \blue{in generic hydrodynamic frame}, which leads to a non-negative
entropy production \blue{in the frame where the first order
conservation laws are restored}.

The order of the fluid division by relaxation time and derivative expansion are different.
In this work, we only calculate the relaxation time up to the first order. At higher orders
of relaxation times, there will still be linear response terms, and the corresponding kinetic
coefficients will become tensors, which lead to more abundant phenomena.
On the other hand,  it is crucial to investigate whether the structure of the covariant linear
response equation is related to the collision model, for example the novel relaxation time model
proposed in \cite{Rocha:2021,Rocha:2022}. We hope to investigate the linear correspondence phenomenon in curved spacetime
in depth in later study.

\appendix

\section{Calculation of particle flow and energy momentum
tensor}\label{Calculation of particle flow and energy momentum tensor}

We work in orthonormal basis $\{(e_{\hat{a}})^{\mu} \}$ obeying
$\eta_{\hat{a} \hat{b}} = g_{\mu \nu}
(e_{\hat{a}})^{\mu}  (e_{\hat{b}})^{\nu}$. Without loss of
generality, we require that $U^{\mu} = c (e_{\hat{0}})^{\mu}$, which implies that the
induced metric can be expressed as $\Delta^{\mu \nu} = \delta^{\hat{i}
\hat{j}}  (e_{\hat{i}})^{\mu}  (e_{\hat{j}})^{\nu}$.

For massive particles, the momentum $p^{\hat{a}} = p^{\mu}
(e^{\hat{a}})_{\mu}$ can be parameterized by mass shell conditions
\begin{equation}
  p^{\hat{a}} = m c (\cosh \vartheta, n^{\hat{i}} \sinh \vartheta), \label{p
  parameter}
\end{equation}
where $n^{\hat{i}} \in S^{d - 1}$ is a spacelike unit vector. Then the
momentum space volume element be represented as
\begin{equation}
  \ensuremath{\boldsymbol{\varpi}} = \frac{(\mathrm{d} p)^d}{| p_{\hat{0}} |}
  = \frac{| \ensuremath{\boldsymbol{p}} |^{d - 1} \mathrm{d} |
  \ensuremath{\boldsymbol{p}} | \mathrm{d} \Omega_{d - 1}}{p^{\hat{0}}} = (m c
  \sinh \vartheta)^{d - 1} \mathrm{d} \vartheta \mathrm{d} \Omega_{d - 1},
\end{equation}
where $\mathrm{d} \Omega_{d - 1}$ is the volume element of the $(d
- 1)$-dimensional unit sphere $S^{d - 1}$. In this way, the
integration in the momentum space can be decomposed into integration over
$\vartheta \in (0, \infty)$ and integration over the unit sphere $S^{d - 1}$.

Here we list some useful integration formulae,
\begin{align}
  \int \mathrm{d} \Omega_{d - 1} & =  \mathcal{A}_{d - 1}, \\
  \int n^{\hat{i}} n^{\hat{j}} \mathrm{d} \Omega_{d - 1} & =  \frac{1}{d}
  \mathcal{A}_{d - 1} \delta^{\hat{i}  \hat{j}}, \\
  \int n^{\hat{i}} n^{\hat{j}} n^{\hat{k}} n^{\hat{l}} \mathrm{d} \Omega_{d -
  1} & =  \frac{3}{(d + 2) d} \mathcal{A}_{d - 1} \delta^{(\hat{i}  \hat{j}}
  \delta^{\hat{k}  \hat{l} )},
\end{align}
\begin{equation}
  \int n^{\hat{i}} \mathrm{d} \Omega_{d - 1} = \int n^{\hat{i}} n^{\hat{j}}
  n^{\hat{k}} \mathrm{d} \Omega_{d - 1} = 0. \label{Sphere integration}
\end{equation}
Using eqs.~\eqref{p parameter}-\eqref{Sphere integration}, we can further calculate
\begin{align}
  \int \ensuremath{\boldsymbol{\varpi}} p^{\mu} f^{(0)}  = &~
  \frac{\mathfrak{g}}{h^d} \mathcal{A}_{d - 1}  (m c)^d J_{d - 1, 1}
  \frac{1}{c} U^{\mu}, \\
  \int \ensuremath{\boldsymbol{\varpi}} p^{\mu} p^{\nu} f^{(0)}  = &~
  \frac{\mathfrak{g}}{h^d} \mathcal{A}_{d - 1}  (m c)^{d + 1}  \left( J_{d -
  1, 2}  \frac{1}{c^2} U^{\mu} U^{\nu} + \frac{1}{d} J_{d + 1, 0} \Delta^{\mu
  \nu} \right), \\
  \int \ensuremath{\boldsymbol{\varpi}}  \frac{1}{p^{\hat{0}}} p^{\mu} p^{\nu}
  f^{(0)}  = &~ \frac{\mathfrak{g}}{h^d} \mathcal{A}_{d - 1}  (m c)^d  \left(
  J_{d - 1, 1}  \frac{1}{c^2} U^{\mu} U^{\nu} + \frac{1}{d} J_{d + 1, - 1}
  \Delta^{\mu \nu} \right), \\
  \int \ensuremath{\boldsymbol{\varpi}}  \frac{1}{p^{\hat{0}}} p^{\mu} p^{\nu}
  p^{\sigma} f^{(0)}  = &~ \frac{\mathfrak{g}}{h^d} \mathcal{A}_{d - 1}  (m
  c)^{d + 1}  \left( J_{d - 1, 2}  \frac{1}{c^3} U^{\mu} U^{\nu} U^{\sigma} +
  \frac{3}{d} J_{d + 1, 0}  \frac{1}{c} U^{(\mu} \Delta^{\nu \sigma)}
  \right), \\
  \int \ensuremath{\boldsymbol{\varpi}}  \frac{1}{p^{\hat{0}}} p^{\mu} p^{\nu}
  p^{\sigma} p^{\rho} f^{(0)}  = &~ \frac{\mathfrak{g}}{h^d} \mathcal{A}_{d -
  1}  (m c)^{d + 2}  \biggl(J_{d - 1, 3}  \frac{1}{c^4} U^{\mu} U^{\nu} U^{\sigma}
  U^{\rho} \nonumber\\
  &~  + \frac{6}{d} J_{d + 1, 1}  \frac{1}{c^2} U^{(\mu} U^{\nu}
  \Delta^{\sigma \rho)} + \frac{3}{(d + 2) d} J_{d + 3, - 1} \Delta^{(\mu \nu}
  \Delta^{\sigma \rho)}\biggl).
\end{align}
The following calculations are then straightforward,
\begin{align}
  N^{(1) \mu}  = &~ c \int \ensuremath{\boldsymbol{\varpi}} p^{\mu} f^{(1)}
  \nonumber\\
   = & - c^3 \tau \int \ensuremath{\boldsymbol{\varpi}}
  \frac{1}{\varepsilon} p^{\mu}  (- p^{\nu} p^{\sigma} \nabla_{\nu}
  \mathcal{B}_{\sigma} + p^{\nu} \nabla_{\nu} \alpha)  \frac{\partial
  f^{(0)}}{\partial \alpha} \nonumber\\
   = & - c^2 \tau \left( - \nabla_{\nu} \mathcal{B}_{\sigma}
  \frac{\partial}{\partial \alpha} \int \ensuremath{\boldsymbol{\varpi}}
  \frac{1}{p^{\hat{0}}} p^{\mu} p^{\nu} p^{\sigma} f^{(0)} + \nabla_{\nu}
  \alpha \frac{\partial}{\partial \alpha} \int
  \ensuremath{\boldsymbol{\varpi}}  \frac{1}{p^{\hat{0}}} p^{\mu} p^{\nu}
  f^{(0)} \right) \nonumber\\
   = & - c^2 \tau \frac{\mathfrak{g}}{h^d}  (m c)^d \mathcal{A}_{d - 1}
  \biggl[\left( \frac{\partial J_{d - 1, 1}}{\partial \alpha}  \frac{1}{c^2}
  U^{\nu} \nabla_{\nu} \alpha - m \frac{\partial J_{d - 1, 2}}{\partial
  \alpha}  \frac{1}{c^2} U^{\nu} U^{\sigma} \nabla_{\nu} \mathcal{B}_{\sigma}
  - m \frac{1}{d}  \frac{\partial J_{d + 1, 0} }{\partial \alpha} \Delta^{\nu
  \sigma} \nabla_{\nu} \mathcal{B}_{\sigma} \right) U^{\mu} \nonumber\\
  &   + \frac{1}{d} \left( - 2 m \frac{\partial J_{d + 1, 0} }{\partial
  \alpha} U^{\sigma} \nabla_{(\sigma} \mathcal{B}_{\nu)} + \frac{\partial J_{d
  + 1, - 1}}{\partial \alpha} \nabla_{\nu} \alpha \right) \Delta^{\mu \nu}\biggl]
  \nonumber\\
   = & - \tau \frac{\mathfrak{g}}{h^d}  (m c)^d \mathcal{A}_{d - 1}
  \biggl[\left( \frac{\partial J_{d - 1, 1}}{\partial \alpha} U^{\nu}
  \nabla_{\nu} \alpha + m c^2  \frac{\partial J_{d - 1, 2}}{\partial \alpha}
  U^{\nu} \nabla_{\nu} \beta - m c^2  \frac{1}{d}  \frac{\partial J_{d + 1, 0}
  }{\partial \alpha} \beta \nabla_{\nu} U^{\nu} \right) U^{\mu} \nonumber\\
  &   + \frac{c^2}{d} \left( \frac{\partial J_{d + 1, - 1}}{\partial \alpha}
  \nabla_{\nu} \alpha + m c^2  \frac{\partial J_{d + 1, 0} }{\partial \alpha}
  \left( \nabla_{\nu} \beta - \frac{1}{c^2} \beta U^{\rho} \nabla_{\rho}
  U_{\nu} \right) \right) \Delta^{\mu \nu}\biggl],
\end{align}
\begin{align}
  T^{(1) \mu \nu}  = &~ c \int \ensuremath{\boldsymbol{\varpi}} p^{\mu}
  p^{\nu} f^{(1)} \nonumber\\
   = & - c^3 \tau \int \ensuremath{\boldsymbol{\varpi}}
  \frac{1}{\varepsilon} p^{\mu} p^{\nu}  (- p^{\rho} p^{\sigma} \nabla_{\rho}
  \mathcal{B}_{\sigma} + p^{\sigma} \nabla_{\sigma} \alpha)  \frac{\partial
  f^{(0)}}{\partial \alpha} \nonumber\\
   = & - c^2 \tau \left( - \nabla_{\rho} \mathcal{B}_{\sigma}
  \frac{\partial}{\partial \alpha} \int \ensuremath{\boldsymbol{\varpi}}
  \frac{1}{p^{\hat{0}}} p^{\mu} p^{\nu} p^{\rho} p^{\sigma} f^{(0)} +
  \nabla_{\sigma} \alpha \frac{\partial}{\partial \alpha} \int
  \ensuremath{\boldsymbol{\varpi}}  \frac{1}{p^{\hat{0}}} p^{\mu} p^{\nu}
  p^{\sigma} f^{(0)} \right) \nonumber\\
   = & - c \tau \frac{\mathfrak{g}}{h^d} \mathcal{A}_{d - 1}  (m c)^{d + 1}
  \biggl[\frac{1}{c^2} \left( - m c^2  \frac{\partial J_{d - 1, 3} }{\partial
  \alpha}  \frac{1}{c^2} U^{\sigma} U^{\rho} \nabla_{\rho}
  \mathcal{B}_{\sigma} - m c^2  \frac{1}{d}  \frac{\partial J_{d + 1,
  1}}{\partial \alpha} \nabla_{\rho} \mathcal{B}_{\sigma} \Delta^{\sigma \rho}
  + \frac{\partial J_{d - 1, 2}}{\partial \alpha} U^{\sigma} \nabla_{\sigma}
  \alpha \right) U^{\mu} U^{\nu} \nonumber\\
  &   + \frac{1}{d} \left( - 2 m c^2  \frac{\partial J_{d + 1, 1}}{\partial
  \alpha}  \frac{1}{c^2} U^{\sigma} \nabla_{(\rho} \mathcal{B}_{\sigma)} +
  \frac{\partial J_{d + 1, 0}}{\partial \alpha} \nabla_{\rho} \alpha \right) 2
  U^{(\mu} \Delta^{\nu) \rho} \nonumber\\
  &   + \frac{1}{d} (- m c^2  \frac{\partial J_{d + 1, 1}}{\partial \alpha}
  \frac{1}{c^2} \nabla_{\rho} \mathcal{B}_{\sigma} U^{\sigma} U^{\rho} - m c^2
  \frac{1}{d}  \frac{\partial J_{d + 3, - 1}}{\partial \alpha} \nabla_{\rho}
  \mathcal{B}_{\sigma} \Delta^{\sigma \rho} + \frac{\partial J_{d + 1,
  0}}{\partial \alpha} U^{\sigma} \nabla_{\sigma} \alpha) \Delta^{\mu \nu}
  \nonumber\\
  &   - m c^2  \frac{2}{(d + 2) d}  \frac{\partial J_{d + 3, - 1}}{\partial
  \alpha}  \left( \nabla_{(\rho} \mathcal{B}_{\sigma)} \Delta^{\mu \sigma}
  \Delta^{\nu \rho} - \frac{1}{d} \nabla_{\rho} \mathcal{B}_{\sigma}
  \Delta^{\sigma \rho} \Delta^{\mu \nu} \right)\biggl] \nonumber\\
   = & - c \tau \frac{\mathfrak{g}}{h^d} \mathcal{A}_{d - 1}  (m c)^{d + 1}
  \biggl[\frac{1}{c^2} \left( \frac{\partial J_{d - 1, 2}}{\partial \alpha}
  U^{\sigma} \nabla_{\sigma} \alpha + m c^2  \frac{\partial J_{d - 1, 3}
  }{\partial \alpha} U^{\sigma} \nabla_{\sigma} \beta - m c^2  \frac{1}{d}
  \frac{\partial J_{d + 1, 1}}{\partial \alpha} \beta \nabla_{\rho} U^{\rho}
  \right) U^{\mu} U^{\nu} \nonumber\\
  &   + \frac{1}{d} \left( \frac{\partial J_{d + 1, 0}}{\partial \alpha}
  \nabla_{\rho} \alpha + m c^2  \frac{\partial J_{d + 1, 1}}{\partial \alpha}
  \left( \nabla_{\rho} \beta - \frac{1}{c^2} \beta U^{\sigma} \nabla_{\sigma}
  U_{\rho} \right) \right) (\Delta^{\rho \nu} U^{\mu} + \Delta^{\rho \mu}
  U^{\nu}) \nonumber\\
  &   + \frac{1}{d} \left( \frac{\partial J_{d + 1, 0}}{\partial \alpha}
  U^{\rho} \nabla_{\rho} \alpha + m c^2  \frac{\partial J_{d + 1, 1}}{\partial
  \alpha} U^{\rho} \nabla_{\rho} \beta - m c^2  \frac{1}{d}  \frac{\partial
  J_{d + 3, - 1}}{\partial \alpha} \beta \nabla_{\rho} U^{\rho} \right)
  \Delta^{\mu \nu} \nonumber\\
  &   - m c^2  \frac{2}{(d + 2) d}  \frac{\partial J_{d + 3, - 1}}{\partial
  \alpha} \beta \left( \Delta^{\rho \mu} \Delta^{\sigma \nu} \nabla_{(\rho }
  U_{ \sigma)} - \frac{1}{d} \nabla_{\rho} U^{\rho} \Delta^{\mu \nu} \right)
  \biggl].
\end{align}

\section*{Acknowledgement}

This work is supported by the National Natural Science Foundation of China under the grant
No. 12275138 and by the Hebei NSF under the Grant No. A2021205037.

\section*{Data Availability Statement}

This work is purely theoretical and contains only analytic analysis.
Hence there is no associated numeric data.

\section*{Declaration of competing interest}

The authors declare no competing interest.

\providecommand{\href}[2]{#2}\begingroup\raggedright
\endgroup

\end{document}